\documentclass[12pt,epsf]{article}

\usepackage{graphicx}

\def\lambdabar{\lambda\raise0.4ex\hbox{\kern-0.5em\hbox{--}}\ }
\def\lesssim{{\lower0.5ex\hbox{$\stackrel{<}{\sim}$}}}

\begin{document}

\pagenumbering{roman}

\vspace*{1 cm}

\noindent
\begin{center}
{\Large \bf Considerations on the Diffraction Limitations to the Spatial
 Resolution of Optical Transition Radiation}
\end{center}

\vspace{0.8 cm}

\noindent
X.Artru$^{a}$, R.Chehab$^{b}$, K.Honkavaara$^{b,c,*}$, A.Variola$^{b,d}$ 
\vspace{0.4 cm}

\noindent   
$ ^{a}$ Institut de Physique Nucl\'eaire de Lyon IN2P3-CNRS and 
Universit\'e Claude Bernard, F-69622 Villeurbanne Cedex, France
\newline  
$ ^{b}$ Laboratoire de l'Acc\'el\'erateur Lin\'eaire IN2P3-CNRS,
Universit\'e de Paris-Sud, B.P.34, F-91898 Orsay Cedex, France
\newline
$ ^{c}$ Helsinki Institute of Physics, P.O.Box 9, FIN-00014 University 
of Helsinki, Finland
\newline
$ ^{d}$ presently at CERN
\newline

\vspace{0.2 cm}

\noindent
$^{*}$ Corresponding author. Fax.+33 1 69071499, 
e-mail: khonkava@lalcls.in2p3.fr

\vspace{0.5 cm}

\section*{Abstract}

The interest in using optical transition radiation (OTR) in high energy
(multiGeV) beam diagnostics has motivated theoretical and experimental
investigations on the limitations brought by diffraction on the attainable 
resolution. This paper presents calculations of the diffraction effects in 
an optical set-up using OTR. The OTR diffraction pattern in a telescopic 
system is calculated taking into account the radial polarization of OTR. 
The obtained diffraction pattern is compared to the patterns obtained by 
other authors and the effects of different parameters on the shape and on 
the size of the OTR diffraction pattern are studied. The major role played
by the radial polarization on the shape of the diffraction pattern is
outlined. An alternative method to calculate the OTR diffraction pattern
is also sketched.

\vspace{0.5 cm}
\noindent
{\it Keywords:} optical transition radiation, spatial resolution, 
diffraction, polarization

\pagebreak

\pagenumbering{arabic}

\section{Introduction}

Optical transition radiation (OTR) provides an attractive method for 
diagnostics of charged particle beams and it has been used for
instance for electron beam diagnostics in the keV-MeV energy region. 
There have been, however, statements that the geometrical resolution of 
OTR might deteriorate drastically at high energies due to the diffraction
phenomenon \cite{mcd}. During the last years there have been 
several studies concerning the resolution of OTR (see Refs.\cite{je} -
\cite{mic}) and this paper extends these investigations concentrating in
the optical diffraction of OTR in a telescopic system.

In order to study the resolution of the optical transition radiation, we shall
calculate the diffraction pattern of OTR on the image plane  of a telescope, 
which is situated in the direction of specular reflection of the incident 
particle (i.e.\ only the case of backward OTR is considered). Naturally, the 
results are valid on the image plane of any kind of imaging system. Scalar 
diffraction theory (see, for example, Ref.\cite{bw} or Ref.\cite{ll}) used by 
D.W.Rule and R.B.Fiorito in Refs.\cite{rf}-\cite{rf2} does not take into 
account the polarization of the field. Precisely, in the case of OTR the 
polarization is important, since the polarization of OTR is not uniform, but 
radial (the electric field is in a plane containing the wave vector and the 
direction of the specular reflection). This can be taken into account 
by considering separately the horizontal and the vertical field components. 
The method is similar to that used by A.Hofmann and F.M\'eot in Ref.\cite{hof}
for synchrotron radiation. Our treatment yields a satisfactory description of 
the diffraction phenomenon and provides a relatively simple, clear and 
straightforward method to compute the transition radiation diffraction pattern.

After recalling some basic characteristics of the optical transition
radiation and of the scalar diffraction theory we shall consider 
the diffraction effects of diaphragms in a telescope and use the obtained
expression in the particular case of OTR. The diffraction pattern of 
OTR will thus allow us to study the influence of different parameters
on its shape and size. Our result will also be compared with others using the 
same hypothesis (polarized character of OTR) \cite{leb}, \cite{mic}. 
A comparison of the OTR diffraction pattern with the well known standard 
diffraction pattern and with the "scalar" diffraction pattern similar to that 
obtained by D.W.Rule and R.B.Fiorito \cite{rf}-\cite{rf2} will be presented. 
As OTR is radially polarized, a comparison with isotropic radiation, radially 
polarized, will allow us to precise the respective contribution of the 
non-constant angular distribution of OTR. A more theoretical treatment will 
also be given for the OTR diffraction.

\section{Recalls}

Before calculating the diffraction pattern of OTR in a telescopic system,
we shall recall some basic characteristics of OTR and of the scalar diffraction
theory.

\subsection{Optical transition radiation}

Transition radiation is emitted when a charged particle crosses 
a boundary between two media of different optical properties.
The emission occurs both into the forward and backward hemispheres
with respect to the separating surface. Here, we shall consider the case
of a single boundary between a metal and vacuum. Due to metal opacity,
only forward (resp.\ backward) OTR is observed when the electron moves from
metal to vacuum (resp.\ vacuum to metal). If the surface is perfectly 
reflecting ($r = r_{\parallel} = r_{\perp} = -1$), the angular distribution
is approximately given (in Gaussian units) by (see, for example, Ref.\cite{lw}):

\vspace*{0.5cm}
\begin{equation} \label{lw1}
I(\theta) = \frac{\mathrm{d}^{2} W}{\mathrm{d} \omega \mathrm{d} \Omega} =
\frac{e^{2}}{\pi^{2}c} \left(\frac{\theta}
{\gamma^{-2}+\theta^{2}}\right)^2
\end{equation}
\vspace*{0.5cm}

\noindent
where $\theta$ is the angle with respect to the electron velocity (forward 
OTR) or to the direction of the specular reflection of that velocity 
(backward OTR). $\gamma$ is the Lorentz factor of the electron, and
Eq.(\ref{lw1}) is valid for $\gamma \gg 1$ and $\theta \ll 1$.
From now on we will consider the case of backward OTR (Fig.1). 

The emitted electric field has two polarization components: one in
the plane of observation ($\hat z_{s} \hat n$ -plane in Fig.1) and the 
other one in the plane perpendicular to that. In the transverse plane 
perpendicular to the direction of specular reflection ($\hat x \hat y$ 
-plane in Fig.1), the electric field is radially polarized and in that 
plane it can be decomposed into horizontal ($\hat x$-direction) and 
vertical ($\hat y$-direction) components. 
\vspace*{0.5cm}

\begin{figure}[h]
\centering
\scalebox{0.7}{\includegraphics*{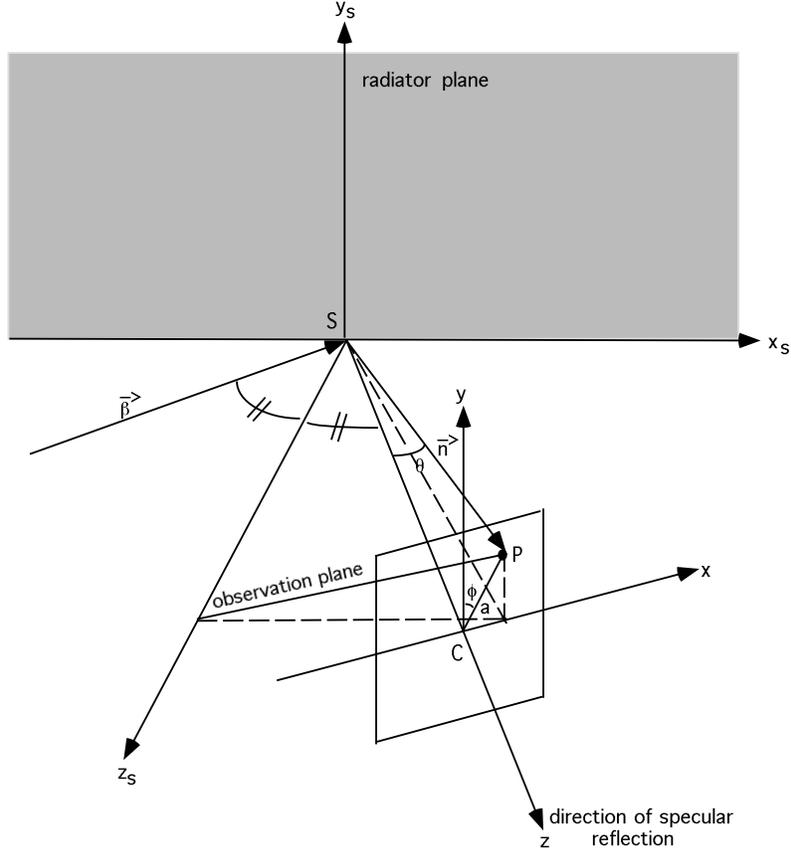}}
\vspace*{0.5cm}
\caption{Definition of coordinates and planes.}
\vspace*{0.5cm}
\end{figure}

\subsection{Transformation of image fields by an optical system
in the scalar wave diffraction theory}

Let us first consider a wave of frequency $\omega = (c/n)k$ propagating 
between two planes $\Pi$ and $\Pi'$ without any lenses between them. In the 
scalar wave theory, with the approximation of Gaussian optics, the amplitude 
$\psi(P')$ in the plane $\Pi'$ is related to the amplitude $\psi_{0}(P)$ in 
the plane $\Pi$ by

\begin{equation}\label{w1}
\psi(P') = -\frac{i}{\lambda} \int_{\Pi} \psi_{0}(P) \frac{e^{ikR}}{R}
\mathrm{d}S
\end{equation}

\noindent
where $R$ is the distance between points $P$ and $P'$. The time-dependent 
factor $e^{-i\omega t}$ has been factored out both in $\psi$ and $\psi_{0}$.
We can treat the ($1/R$)-factor as a constant (in the Gaussian optics 
approximation) and write 

\begin{equation}\label{w2}
\psi(P') = A \int_{\Pi} \psi_{0}(P) e^{ikR} \mathrm{d}S
\end{equation}

\noindent 
where $A = i/(\lambda L)$ and $L$ is the distance between the two planes.

Let us now consider the case where one or several "non-diaphragmed"
lenses are inserted between the planes $\Pi$ and $\Pi'$. By "non-diaphragmed" 
lens, we mean a lens with an aperture much larger than the transverse size of 
the optical wave packet. Eq.(\ref{w2}) can be generalized as\,\footnote{
Eq.(\ref{sdt}) is applicable provided that the planes $\Pi$ and
$\Pi'$ are not conjugate.}

\begin{equation}\label{sdt}
\psi(P') = A \int_{\Pi} \psi_{0}(P) e^{ik \mathcal{L}(P,P')} \mathrm{d}S
\end{equation}

\noindent
where $\mathcal{L}(P,P')$ is the optical distance between points $P$ and 
$P'$, i.e.\ the integral

\begin{equation}\label{l}
\mathcal{L}(P,P') = \int_{P}^{P'} n \mathrm{d} l
\end{equation}

\noindent
along the geometrical optical ray connecting $P$ and $P'$;
$n$ is the refractive index. $A$ is a complex factor, which depends only 
on the location of the planes and which we will not calculate, since we are 
interested only in the shape of the OTR image.

\section{Diffraction effect of diaphragms in a telescope}

In this chapter we consider diffraction effect caused by the 
diaphragms of a telescope, and in the next one we will take 
into account also the special properties of optical transition 
radiation.

We have taken the experimental set-up used in our experiment at Orsay 
\cite{exp} as the geometrical basis for the diffraction calculations. 
In this experiment backward optical transition radiation emitted by a 2 GeV 
electron beam was measured in the direction of the specular reflection. The 
set-up consisted of an OTR radiator, two lenses in {\it a telescopic 
configuration} and a CCD-camera; henceforth the CCD is referred to as a 
screen. The first lens had a focal length of 1 m and a diameter of 8 cm; the 
focal length of the second one was 25 cm and the diameter 14 cm. The first 
lens with a smaller diameter gives the effective aperture limitation of the 
system. The telescope geometry is presented in Fig.2.

\begin{figure}[h]
\centering
\scalebox{0.6}{\includegraphics*{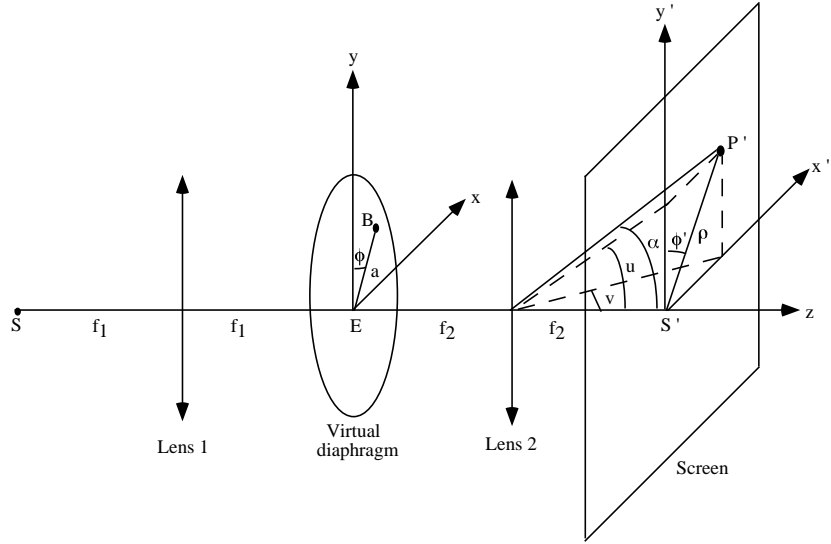}}
\vspace*{-1.5cm}
\caption{Schematic set-up.}
\end{figure}

We will treat the effect of the {\it real} diaphragm (at the first lens) in
a slightly approximate but convenient way, replacing this diaphragm
by a {\it virtual} one, with the same diameter, but located
in the common focal plane between the lenses\,\footnote{The real and the 
virtual diaphragm are in practice equivalent, when $a_{0} \gg \gamma 
\lambdabar$ (the transverse size of the source) and $a_{0}/f_{1}\gg 
\gamma^{-1}$ (the peak angle), where $a_{0}$ and $f_{1}$ are the radius 
and the focal length of the first lens. These two conditions are fulfilled 
in the following calculations.}. In that plane, the spatial coordinates are 
directly related to the angles of the emitted radiation ($a=f_{1}\theta$). 

Our observation point $P'$ is situated in the image plane of the telescope
(see Fig.2) and, according to Eq.(\ref{sdt}), the modulus of the 
field at this point is related to the amplitude $\psi_{0}$ in the common focal
plane by

\begin{equation} \label{sc2}
|\psi(P')| = A |\int_{\epsilon} \psi_{0}(B) e^{ik \mathcal{L}(B,P')} 
\mathrm{d}\epsilon|
\end{equation}

\noindent
where $A$ is a normalization factor. The integration is performed over
the aperture area $\epsilon$ of the virtual diaphragm.

The coordinates of a point $B$ in the virtual diaphragm  
are $x = a \sin \phi$ and $y = a \cos \phi$ (Fig.2). 
In the image plane we use "prime" coordinates: 
$x' = \rho \sin \phi'$ and $y' = \rho \cos \phi'$. The angular directions
$v$ and $u$ in the small angle approximation can be written as 

\begin{eqnarray}
v & = &  \frac{x'}{f_{2}} = \frac{\rho}{f_{2}} 
     \sin \phi' = \alpha \sin \phi'  \label{v} \\
u & = & \frac{y'}{f_{2}} = \frac{\rho}{f_{2}} 
     \cos \phi' = \alpha \cos \phi' \label{u}
\end{eqnarray}

In the phase factor of Eq.(\ref{sc2}) we are interested in the relative
phase difference. All the rays leaving the diaphragm in a particular 
direction are focused by the second lens into the same point of the screen
(see Fig.3). 

\vspace*{0.5cm}
\begin{figure}[h]
\centering
\scalebox{0.6}{\includegraphics*{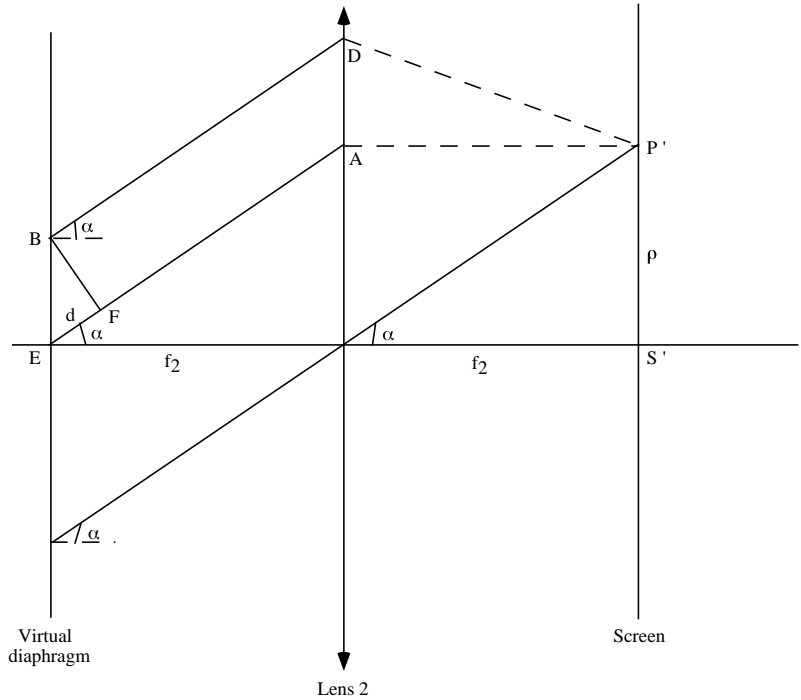}}
\caption{Sketch of rays after the virtual diaphragm.}
\end{figure}

According to the  theorem of Malus \cite{wel}, all the rays
perpendicular to a given surface ($BF$ in Fig.3) have an equal optical 
path length from the surface to the focus point (point $P'$ in Fig.3), and 
thus the optical path difference between the rays $EAP'$ and $BDP'$ is 
$d = |EF|$. This distance is the projection of the vector $\overrightarrow{EB}$
onto the direction of vector $\overrightarrow{EA}$ : $d=|\overrightarrow{EB} 
\cdot (\overrightarrow{EA}/ |\overrightarrow{EA}|)| = vx + uy$, where $x$ and 
$y$ are the coordinates of point $B$ and  $v$ and $u$ are the angular 
directions given by Eq.(\ref{v}) and Eq.(\ref{u}), respectively. The 
corresponding phase difference can now be written as $\delta = -kd = 
-k(vx+uy)$. In polar coordinates the last parenthesis can be written as

\begin{equation}
vx + uy = \frac{x'}{f_{2}}x + \frac{y'}{f_{2}}y 
 = \frac{\rho}{f_{2}} a \cos (\phi - \phi')
\end{equation}

Since OTR is symmetrical about the $z$ -axis, there is no preferred value of 
the angle $\phi'$, and we can select $\phi'=0$ and write the modulus of the 
diffracted amplitude in the point $P'$ on the screen as

\begin{equation} \label{da}
 |\psi(P')| = A |\int_{0}^{2 \pi} \int_{0}^{a_{0}} \psi_{0}(a,\phi)
\mathrm{exp}(-i \frac{2 \pi}{\lambda} \frac{\rho}{f_{2}} a \cos \phi)
a \mathrm{d}a \mathrm{d} \phi |
\end{equation}

\noindent
where $k = 2 \pi/\lambda$ and $\psi_{0}(a,\phi)$ is the amplitude of the wave
in the intermediate focal plane. The integration is made in polar coordinates 
and $a_{0}$ is the radius of the smaller lens (the limiting aperture 
of the system). It should be noticed that Eq.(\ref{da}) is a particular
form of a two-dimensional Fourier transform of the wave amplitude $\psi_{0}$. 
We can therefore say, in agreement with Ref.\cite[Chap.5,\S 5.2.2]{goo},
that the field in the image focal plane of the lens $L_{2}$ is proportional to
the Fourier transform of the field in the object focal plane of $L_{2}$.
In our derivation, this property appears essentially as a consequence of 
the Malus theorem.

\section{Diffraction of OTR in the telescope} \label{chd}

In the calculation of the diffraction pattern of optical transition radiation
(i.e.\ the OTR spot in the image plane of the telescope) we need 
to consider both the amplitude and the polarization of the incident wave. 

The scalar diffraction theory can be used without modification for
the vector case if the direction of the field is the same in each point 
on the diaphragm. The electric field of transition radiation is, however,
radially polarized i.e.\ for every azimuthal angle $\phi$ the field vector 
has a different direction (it is always pointing to the centre of symmetry). 
We may take this characteristic into account by considering separately the 
horizontal and vertical field components. The total intensity is the sum of 
the intensities from these two components. 

When we decompose transition radiation into plane waves with directions
of $\theta$, the amplitude is proportional to $\vec{\theta}/(\gamma^{-2}+
\theta^{2})$, where $\vec{\theta}$ is a two-dimensional vector.
When considering the two polarization components separately, $\vec{\theta}$ 
can be replaced by ($\theta \sin \phi$) for the horizontal and by
($\theta \cos \phi$) for the vertical component\,\footnote{Here the polar
angle $\phi$ is defined with respect to the vertical axis.} (Fig.1).
The phase of these plane waves is precisely zero in the impact point $S$.
 
A plane wave, whose direction of propagation has an angle $\theta$
with respect to the direction of specular reflection and whose azimuthal 
angle is $\phi$, is focused to the point $B = (a,\phi)$ on the virtual 
diaphragm (where $\theta=a/f_{1}$) and the modulus of the electric 
field amplitude in this point is given by 
 
\begin{equation} \label{ef3}
|E_{\omega}(B)| = C'\frac{(a/f_{1})}
{\gamma^{-2} + (a/f_{1})^2}
\end{equation}

\noindent
where $f_{1}$ is the focal length of the first lens  and $C'$ is a constant 
that takes into account the normalization and units.

When considering horizontal and vertical components separately, $|E_{\omega}
(B)|$ has to be multiplied by the factor ($\sin \phi$) or ($\cos \phi$),
respectively, and we can write 

\begin{eqnarray}
\psi_{0h}(a, \phi) & = & |E_{\omega}(B)| \sin \phi \label{adx} \\
\psi_{0v}(a, \phi) & = & |E_{\omega}(B)| \cos \phi \label{ady}
\end{eqnarray}

\noindent 
where $\psi_{0h}(a, \phi)$ refers to the horizontal component and
$\psi_{0v}(a, \phi)$ to the vertical one. We should multiply Eq.(\ref{adx})
and Eq.(\ref{ady}) by a phase factor corresponding to the propagation 
between $S$ and $B$. However, because $E$ and $B$ are on the same wave surface,
the optical paths $SE$ and $SB$ are equal (invoking the Malus theorem), 
and if we forget the constant phase factor, we do not have any extra phase 
factors to add into Eq.(\ref{adx}) and Eq.(\ref{ady}).\footnote{The plane 
wave decomposition of OTR is proportional to the two-dimensional Fourier 
transform of the OTR field at the radiator. Therefore, Eq.(\ref{ef3}), like 
Eq.(\ref{da}), can be considered as an application of 
Ref.\cite[Chap.5,\S 5.2.2]{goo}, the lens being in this case $L_{1}$.}

The total intensity at the point $P'$ is the sum of the intensities from
the horizontal and the vertical components:

\begin{equation}
 I(P') = \vert \mathcal{E}(P') \vert^{2} =
         \vert \mathcal{E}(P') \vert_{h}^{2} 
        + \vert \mathcal{E}(P') \vert_{v}^{2}
\end {equation}
                                                                  
In the case of OTR in a telescope,
$|\mathcal{E}(P')|_{h}$ and $\mathcal{E}(P')|_{v}$ are obtained by
substituting Eq.(\ref{adx}) and Eq.(\ref{ady}) into Eq.(\ref{da}).
By using the expression given by Eq.(\ref{ef3}), we obtain 

\begin{eqnarray}
 \vert \mathcal{E}(P') \vert_{h} & = & A' \vert 
 \int_{0}^{a_{0}} \int_{0}^{2 \pi} \frac{(a/f_{1}) \sin \phi}{\gamma^{-2}
  +(a/f_{1})^2}
  \mathrm{exp}(-i\frac{2 \pi}{\lambda} \frac{\rho}{f_{2}} a \cos 
 \phi) a \mathrm{d}a \mathrm{d} \phi \vert \\
\vert \mathcal{E}(P') \vert_{v} & = & A' \vert 
 \int_{0}^{a_{0}} \int_{0}^{2 \pi} \frac{(a/f_{1}) \cos \phi}{\gamma^{-2}
  +(a/f_{1})^2}
  \mathrm{exp}(-i\frac{2 \pi}{\lambda}\frac{\rho}{f_{2}} a \cos 
 \phi) a \mathrm{d}a \mathrm{d} \phi \vert 
\end{eqnarray}
         
\noindent                              
where $A'$ is a normalization constant. 

The integration over $\phi$ in the horizontal component gives zero, thus
only the vertical component contributes to the total intensity, and we obtain

\begin{equation} \label{dpni}
I(P') = |\mathcal{E}(P')|_{v}^2 = C \vert \int_{0}^{a_{0}}
\frac{a^2}{\gamma^{-2}+(a/f_{1})^2} 
J_{1}(\frac{2 \pi}{\lambda} \frac{\rho}
{f_{2}} a) \mathrm{d} a \vert^2
\end{equation}

\noindent
where $J_{1}$ is the first order Bessel function and $C$ a generic 
normalization constant.

We have defined the angle $\phi$ with respect to the vertical axis. Of course
this angle can as well be defined with respect to the horizontal axis.
In that case the $\sin \phi$ and $\cos \phi$ -factors in Eq.(\ref{adx})
and Eq.(\ref{ady}) are changed with each other and the horizontal
component instead of the vertical one
gives the contribution to the intensity of Eq.(\ref{dpni}).

The integration over $a$ in Eq.(\ref{dpni}) can be performed numerically 
and the result as a function of the radius $\rho$ on the screen is shown in 
Fig.4 for our experimental conditions ($E = 2$ GeV, $f_{1} = 1$ m, $f_{2} = 25$
cm, $a_{0} = 4$ cm, $\theta_{1}=a_{0}/f_{1} = 40$ mrad, $\lambda = 500$ nm). 
This distribution, which represents the diffraction pattern of an OTR source 
taking into account the radial polarization, is shown around the symmetry axis.
Since we are not interested in the absolute intensity, the peak intensity is 
normalized to unity. The magnification of the used telescope is $M=f_{2}/
f_{1}=0.25$; if we use an imaging system with magnification of one, the 
diffraction pattern is naturally four times wider. The FWHM size of the 
pattern in Fig.4 is about 4.5 $\mu$m (FWHM $\approx 18$ $\mu$m, when $M=1$). 
This pattern shape is in full agreement with that of V.A.Lebedev 
obtained in Ref.\cite{leb}. Similar observations concerning this pattern 
shape are presented in Ref.\cite{xav} and Ref.\cite{mic}.
\vspace*{0.4cm}

\begin{figure}[h]
\centering
\scalebox{0.45}{\includegraphics*{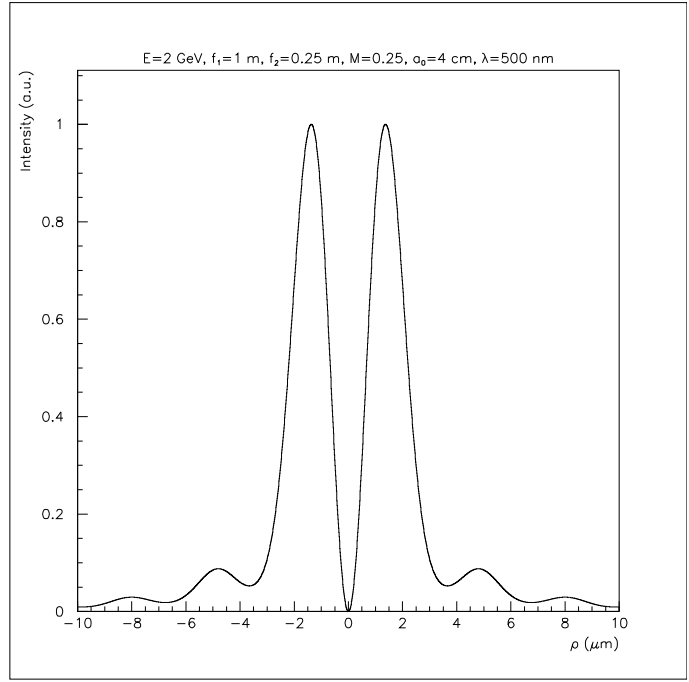}}
\vspace*{0.4cm}
\caption{OTR diffraction pattern (intensity) around the symmetry axis 
on the image plane of a telescope with magnification $M=0.25$ ($E=2$ GeV, 
$f_{1}=1$ m, $f_{2}=25$ cm, $M=f_{2}/f_{1}=0.25$, $a_{0}=4$ cm, 
$\theta_{1}=a_{0}/f_{1}=40$ mrad, $\lambda=500$ nm). In a one-to-one imaging 
system the size of the diffraction pattern is four times larger.}
\vspace*{0.2cm}
\end{figure}

\subsection{Effects of different parameters on the OTR diffraction 
pattern}

Next we study the effects of different parameters on the
OTR diffraction pattern. Since we are only interested in the
size of the pattern, the peak intensities are always scaled to unity. 
In all the figures the magnification of the system is 0.25;
when using a one-to-one imaging system, the patterns are four times
wider.
\newpage

Fig.5 shows the OTR diffraction pattern for different wavelengths. 
We can see, as expected, that the size of the pattern scales
proportionally to the wavelength. The resolution can be improved when
using smaller wavelengths, but if we are out of the optical range
($\lambda \lesssim 350$ nm), we can not use an optical imaging system and
the experimental conditions become more complicated.
\vspace*{0.5cm}
\begin{figure}[h]
\centering
\scalebox{0.45}{\includegraphics*{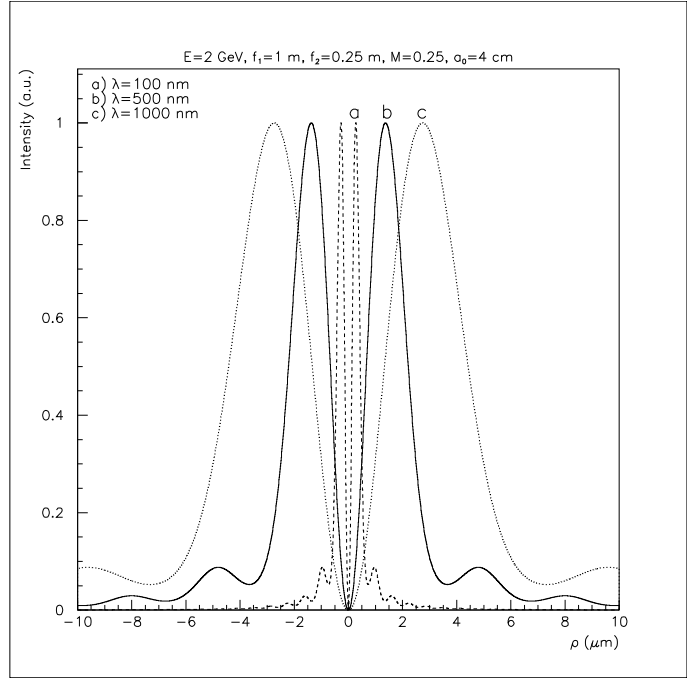}}
\vspace*{0.5cm}
\caption{OTR diffraction pattern for different wavelengths ( a) $\lambda=
100$ nm, b) $\lambda=500$ nm and c) $\lambda = 1$ $\mu$m) on the image plane
of a telescope with magnification $M=0.25$. The used parameters are the same 
as in Fig.4.}
\vspace*{0.5cm}
\end{figure}

In Fig.6 the geometrical size of the aperture ($a_{0}$) is varied.
Naturally, the decrease of the aperture size causes an enlargement
of the diffraction pattern.
\newpage
\begin{figure}[h]
\centering
\scalebox{0.45}{\includegraphics*{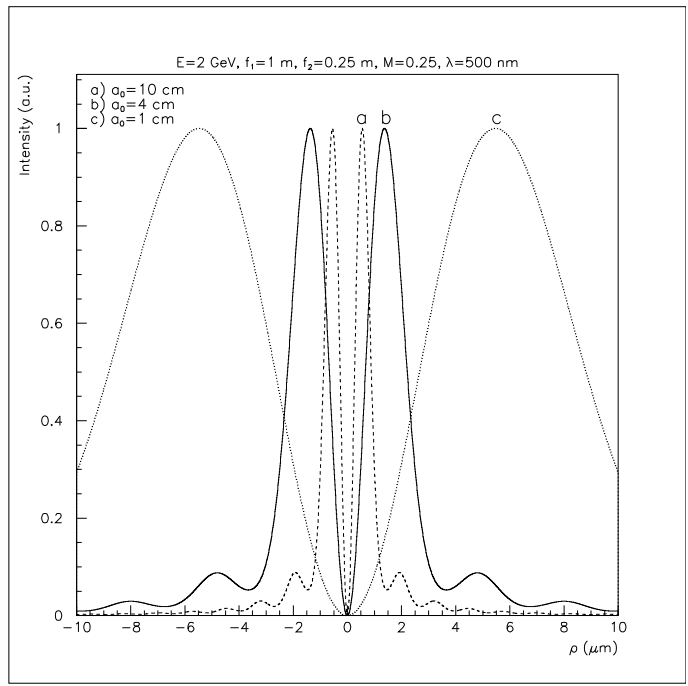}}
\vspace*{0.5cm}
\caption{OTR diffraction pattern for different geometrical aperture sizes\break
(a) $a_{0}=10$ cm, b) $a_{0}=4$ cm and c) $a_{0}=1$ cm) on the image plane
of a telescope with magnification $M=0.25$. The used parameters are the same 
as in Fig.4.}
\vspace*{0.5cm}
\end{figure}

Fig.7 shows OTR diffraction pattern for different energies in
the GeV energy range. The FWHM size of the distribution is independent 
of $\gamma$. The difference is in the tails: the higher is the energy, 
the stronger are the tails.

\begin{figure}[h]
\centering
\scalebox{0.45}{\includegraphics*{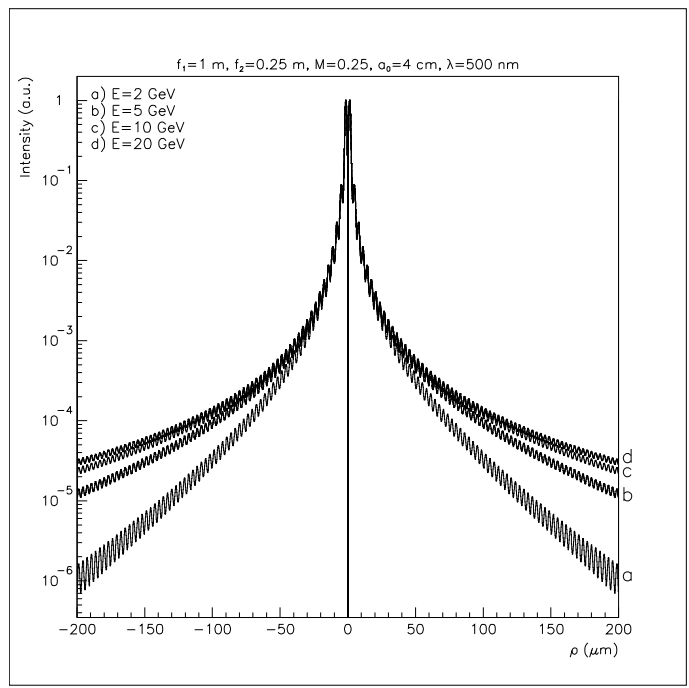}}
\caption{OTR diffraction pattern for different energies (a) $E=2$ GeV,
b) $E=5$ GeV, c) $E=10$ GeV and d) $E=20$ GeV) on the image plane of a 
telescope with magnification $M=0.25$. The used parameters are the same 
as in Fig.4.}
\end{figure}

In Refs.\cite{je},\cite{exp},\cite{bor} and more recently also in 
Refs.\cite{xav} and \cite{mic}, it has been considered a possibility
to use a mask\,\footnote{A "stop" in Ref.\cite{je}.}
to improve the spatial resolution. The effect of a mask can be 
taken into account by introducing into Eq.(\ref{dpni}) an extra pupil function 
representing the cut caused by the mask. It amounts to set the lower
limit of integration in Eq.(\ref{dpni}) to $a_{m}$ instead of zero, where\break
\newpage
\noindent
$a_{m}$ is the radius of the mask\,\footnote{The mask should, in principle,
be put in the common focal plane of lenses $L_{1}$ and $L_{2}$. However, it 
can be put on $L_{1}$, if $a_{m}\gg \gamma \lambdabar$ and $(a_{m}/f_{1})
\gg \gamma^{-1}$ (cf.\ similar conditions than for the real diaphragm).}. 
A mask reduces the tails, as can be seen in Fig.8, where OTR 
diffraction pattern for $E=10$ GeV (M=0.25) has been plotted with 
and without a mask ($a_{mask}=2$ mm). However, it does not affect 
significantly the FWHM size of the pattern.

\begin{figure}[h]
\centering
\scalebox{0.45}{\includegraphics*{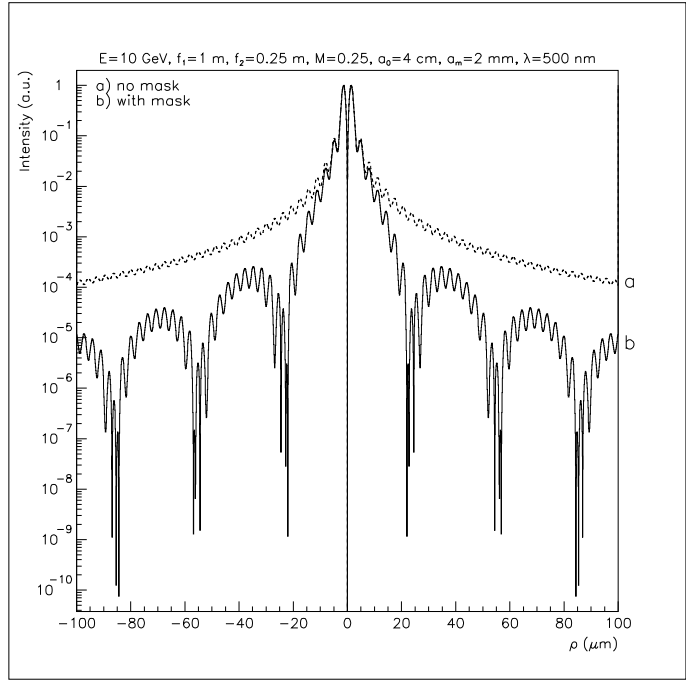}}
\vspace*{0.5cm}
\caption{Effect of a mask with a radius $a_{m}=2$ mm ($a_{m}/a_{0}=
0.05$). $E=\break10$ GeV; the used parameters are the same as in Fig.4.} 
\end{figure}

\subsection{Diffraction of a gaussian emitter}

So far, we have considered OTR emitted by a single electron.
Diffraction of OTR emitted by a gaussian beam can be treated by  
convoluting on the image plane the OTR diffraction pattern 
and a gaussian distribution, which is the image of the beam distribution.
In a general form, this is a two-dimensional convolution:

\begin{equation}
I_{\mathrm{conv}}(x,y) =  \int \int I(x-x_{1},y-y_{1})\mathcal{O}(x_{1},y_{1})
\mathrm{d}x_{1} \mathrm{d}y_{1}
\end{equation}

\noindent
where $I(x,y)$ is the OTR diffraction pattern (Eq.(\ref{dpni})) in cartesian
coordinates ($\rho = \sqrt{x^2+y^2}$) and $\mathcal{O}(x,y)$ the image of
the beam profile:

\begin{equation}\label{gd}
\mathcal{O}(x,y) = \frac{1}{\sqrt{2\pi}\sigma_{ix}} \mathrm{exp}(-\frac{x^2}
{2\sigma_{ix}^2}) \frac{1}{\sqrt{2\pi}\sigma_{iy}} \mathrm{exp}(-\frac{y^2}
{2\sigma_{iy}^2})
\end{equation}

\noindent
where $\sigma_{ix}$ and $\sigma_{iy}$ are the horizontal and vertical
rms sizes of the gaussian beam image.

\section{Comparison with standard diffraction}

Let us calculate for comparison the diffraction pattern of an ideal isotropic 
point source (the standard diffraction pattern) in a telescope. In that case 
$\psi_{0}(a, \phi) = {\rm constant}$. By substituting this into Eq.(\ref{da}) 
and squaring, we obtain

\begin{equation}
 I(P') = \mathrm{const}*| \int_{0}^{2 \pi}
 \int_{0}^{a_{0}} \mathrm{exp}(-i \frac{2 \pi}{\lambda} \frac{\rho}
{f_{2}} a \cos \phi) a \mathrm{d} a \mathrm{d} \phi|^2
\end{equation}

Integration over $\phi$ and $a$ gives

\begin{equation} \label{st}
I(P') = C \vert \frac{J_{1}\left(\frac{2 \pi}{\lambda}\frac{\rho}{f_{2}} 
a_{0}\right)}{\left(\frac{2 \pi}{\lambda}\frac{\rho}{f_{2}} a_{0} \right)} 
\vert^2 
\end{equation}

\noindent
where $J_{1}$ is the first order Bessel function and
$C$ is a generic normalization constant.

In Fig.9 the standard diffraction pattern given by Eq.(\ref{st}) (curve a)
is compared with the OTR diffraction pattern given by Eq.(\ref{dpni})
(curve c) in our experimental conditions. The peak intensities are 
both normalized to unity. We can see that the OTR diffraction pattern is 
wider (the FWHM size is about 2.7 times that of the standard diffraction
pattern) and has a zero in the center.

\section{Diffraction of "scalar OTR"}

If we do not take into account the radial polarization of OTR, but only
the angular distribution of it, we can use the right hand side of 
Eq.(\ref{ef3}) as the algebraic amplitude. By substituting it into
Eq.(\ref{da}) we obtain

\begin{equation}
|\mathcal{E}(P')| = A' |\int_{0}^{2 \pi}
\int_{0}^{a_{0}} \frac{(a/f_{1})}{\gamma^{-2}+(a/f_{1})^2}
\mathrm{exp}(-i \frac{2 \pi}{\lambda} \frac{\rho}{f_{2}} a \cos \phi)
a \mathrm{d} a \mathrm{d} \phi |
\end{equation}

After integrating over $\phi$ we have

\begin{equation} \label{dwp}
I(P') = |\mathcal{E}(P')|^2 = C |\int_{0}^{a_{0}}  \frac{a^2}
{\gamma^{-2}+(a/f_{1})^2} J_{0} \left(\frac{2 \pi}
{\lambda} \frac{\rho}{f_{2}} a \right) \mathrm{d} a |^2
\end{equation}

\noindent
where  $J_{0}$ is the zero order Bessel function and $C$ is a generic 
normalization constant.

The integration over $a$ can again be performed  numerically and the resulting
diffraction pattern is plotted in Fig.9 (curve b). The peak intensity is
again normalized to unity. This pattern is similar to the pattern obtained by 
D.W.Rule and R.B.Fiorito in Refs.\cite{rf}-\cite{rf2}. 
The FWHM size of this "scalar OTR" diffraction pattern is by a factor
$\sim$ 1.2 wider than the FWHM size of the standard diffraction pattern; the 
FWHM size of the "vector OTR" pattern given by Eq.(\ref{dpni}) is by a factor 
of $\sim$  2.2 wider than the FWHM size of the "scalar" one (Eq.(\ref{dwp})).
\vspace*{0.5cm}

\begin{figure}[h]
\centering
\scalebox{0.45}{\includegraphics*{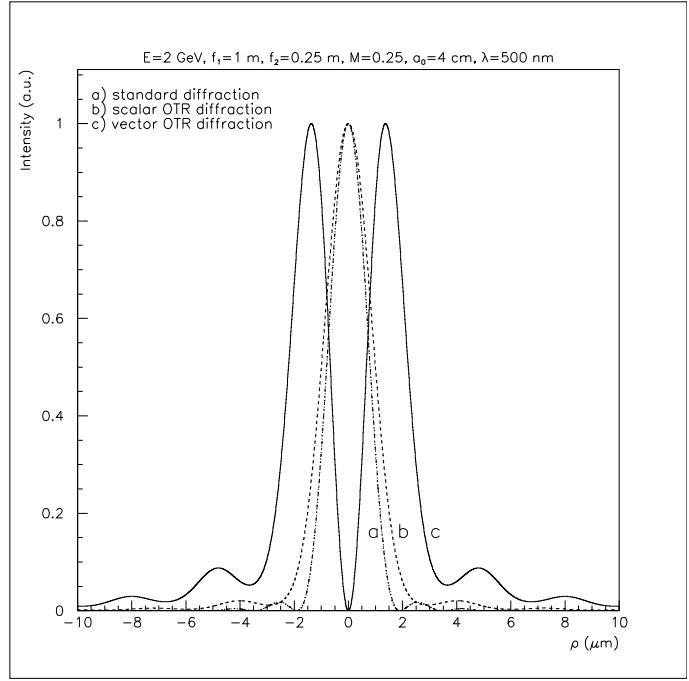}}
\vspace*{0.5cm}
\caption{Standard diffraction pattern given by Eq.(\ref{st}) (curve a),
"vector" OTR diffraction pattern given by Eq.(\ref{dpni}) (curve c) and 
"scalar" OTR diffraction pattern given by Eq.(\ref{dwp}) (curve b) on the 
image plane of a telescope with magnification  $M=0.25$. 
The used parameters are the same as in Fig.4.}
\end{figure}

\section{Importance of the radial polarization in the diffraction
phenomenon}

For a field which is invariant by rotation about the direction of the
specular reflection, the amplitude distribution and the 
polarization of the field can be described by separate functions $\mathcal{A}$
and $\mathcal{F}$, respectively. According to Eq.(\ref{da})
we can write

\begin{equation}
|\mathcal{E}(P')|_{i}  =  A |\int_{0}^{2 \pi}
\int_{0}^{a_{0}} \mathcal{A}(a) \mathcal{F}_{i}(\phi)
\mathrm{exp}(-i \frac{2 \pi}{\lambda} \frac{\rho}{f_{2}} a \cos \phi) 
a \mathrm{d} a \mathrm{d} \phi |  \label{pl}
\end{equation}

\noindent
where index $i$ refers to the horizontal or to the vertical component.
The total intensity is the sum of the intensities from different
components: $I(P') = \sum_{i}|\mathcal{E}(P')|_{i}^2$

When the field is radially polarized, the polarization function is 
$\mathcal{F}_{h}(\phi) = \sin \phi$ for the horizontal component and 
$\mathcal{F}_{v}(\phi) = \cos \phi$ for the vertical one. The angle $\phi$ 
is again defined with respect to the vertical axis.
Let us consider a hypothetical case in which the field is constant
in the amplitude : $\mathcal{A}(a) = \mathrm{constant}$.
By substituting these into Eq.(\ref{pl}), we obtain

\begin{eqnarray}
|\mathcal{E}(P')|_{h} & = & \mathrm{const}*|\int_{0}^{2 \pi} \int_{0}^{a_{0}}
 \sin \phi \mathrm{exp}(-i \frac{2 \pi}{\lambda} \frac{\rho}{f_{2}} a \cos 
\phi)a \mathrm{d}a \mathrm{d} \phi| \\
|\mathcal{E}(P')|_{v} & = & \mathrm{const}*|\int_{0}^{2 \pi} \int_{0}^{a_{0}} 
 \cos \phi \mathrm{exp}(-i \frac{2 \pi}{\lambda} \frac{\rho}{f_{2}} a \cos 
\phi) a \mathrm{d} a \mathrm{d} \phi| 
\end{eqnarray}

The integration over $\phi$ in the horizontal component gives again zero 
and we obtain for the total intensity 

\begin{equation} \label{pdp}
 I(P') = |\mathcal{E}(P')|_{v}^{2} = C |\int_{0}^{a_{0}} a
J_{1} \left(\frac{2 \pi}{\lambda} \frac{\rho}{f_{2}} a \right)
da|^2
\end{equation}

\noindent
where $C$ is a generic normalization constant.

Eq.(\ref{pdp}) is plotted in Fig.10 (curve a) together with the OTR diffraction pattern (curve b). The peak intensities are both normalized to unity. It is 
important to understand that the peculiar shape of the OTR diffraction 
pattern in the central region with a zero in the center is essentially 
determined by the radial polarization. The non-constant angular distribution 
of OTR only widens the pattern a little: the FWHM value is wider by a factor 
of $\sim$ 1.2, when the OTR angular distribution is taken into account.

\begin{figure}[h]
\centering
\scalebox{0.45}{\includegraphics*{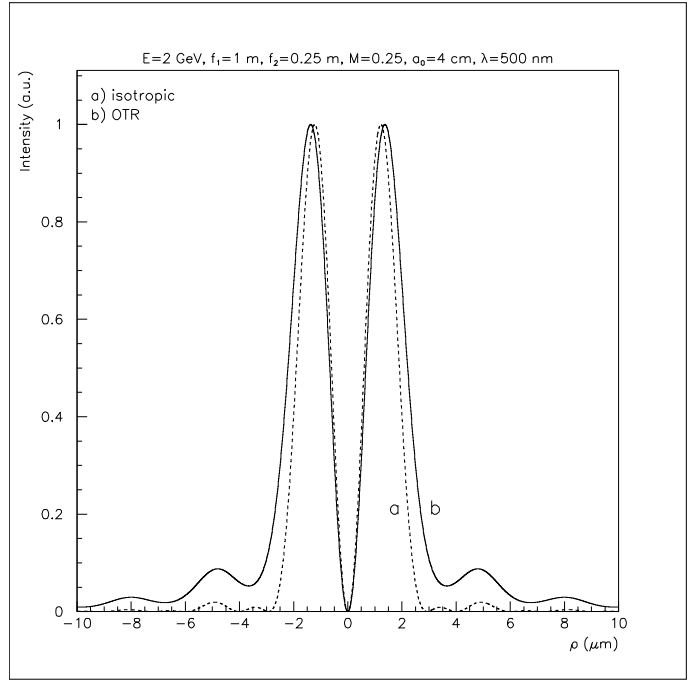}}
\caption{Diffraction pattern on the image plane of a telescope with
magnification M=0.25 given by radial polarization:
a) isotropic angular distribution (Eq.(\ref{pdp})) and b)
OTR angular distribution (Eq.(\ref{dpni})). The used parameters 
are the same as in Fig.4.}
\end{figure}

\section{Another treatment of OTR diffraction}

Diffraction of OTR can also be studied from a more theoretical
point of view. A detailed treatment of this kind is presented
elsewhere \cite{xav} and here we only shortly show that we can obtain,
using this method, the same expression for diffraction pattern as  
obtained in paragraph \ref{chd}.

The angular distribution of transition radiation in natural units 
($c=\hbar=\epsilon_{0}=1$, $e^2/4 \pi = \alpha = 1/137$) can be written as

\begin{equation} \label{adt}
 I(\omega,\theta) = \frac{\mathrm{d}^2I}{\mathrm{d}\omega \mathrm{d}
\Omega} = \omega \frac{\mathrm{d}N}{\mathrm{d}\omega \mathrm{d}\Omega}
\simeq \frac{\alpha}{\pi^2} \left(\frac{\theta}{\gamma^{-2}+\theta^2}
\right)^2
\end{equation}

In the case of forward radiation Eq.(\ref{adt}) is the spectrum emitted by a
suddenly accelerated electron and in the backward case the spectrum emitted
by a suddenly stopped "image positron". The radiation field (in the far-field
region) can be decomposed in plane waves

\begin{equation} \label{pw}
\mathbf{E}(t,\mathbf{\vec{r}}) = \int \frac{\mathrm{d}^3\mathbf{\vec{k}}}
{(2 \pi)^3} \mathbf{\tilde{E}}(\mathbf{\vec{k}}) e^{i\mathbf{\vec{k}} \cdot 
\mathbf{\vec{r}}-i|\mathbf{\vec{k}}|t}
\end{equation}

with

\begin{equation}\label{ef}
\mathbf{\tilde{E}}(\mathbf{\vec{k}}) \simeq ie \frac{\mathbf{\vec{q}}}
{\mathbf{\vec{q}}^2+\gamma^{-2}k_{L}^{2}}
\end{equation}

\noindent
where $\mathbf{\vec{q}}$ and $\mathbf{\vec{k_{L}}}$ are the transverse
and the longitudinal components of the wave vector $\mathbf{\vec{k}}$,
respectively.

The impact parameter profile\,\footnote{Impact parameter $\mathbf{\vec{b}}$ is
defined as the transverse distance of the photon to the incident particle and
$b = |\mathbf{\vec{b}}|$ is related to the radial coordinate $\rho$ by 
$b=\rho/M$.} is related to the $\mathbf{\vec{q}}$ -Fourier transform 
of $\mathbf{\tilde{E}}$:

\begin{eqnarray}
I(b) & \equiv & \frac{\omega}{\mathrm{d}\omega} \frac{\mathrm{d}N}
{\mathrm{d}^2 \mathbf{\vec{b}}} \simeq \frac{1}{\pi} |\mathbf{E}(\omega,
\mathbf{\vec{b}}|^2 \nonumber \\
& = & 4 \alpha \vert \int  \frac{\mathrm{d}^2\mathbf{\vec{q}}}{(2 \pi)^2}
\frac{\mathbf{\vec{q}}}{\mathbf{\vec{q}}^2+q_{0}^2} f(q)
e^{i \mathbf{\vec{q}} \cdot \mathbf{\vec{b}}} \vert^2  \label{ib1}
\end{eqnarray}

\noindent
where $f(q)$ is a cut-off function:

\begin{equation}\label{fq}
f(q) = \Theta(q-q_{m}) \Theta(q_{1}-q)
\end{equation}

The parameters $q_{0}$, $q_{1}$ and $q_{m}$ are defined as

\begin{eqnarray}
q_{0} & = & \gamma^{-1} k_{L} \simeq \gamma^{-1} \omega \label{q0} \\
q_{1} & = & \theta_{1} \omega \label{q1} \\
q_{m} & = & \theta_{m} \omega \label{qm}
\end{eqnarray}

\noindent
where $\theta_{1}$ is the upper cut-off angle determined by some
diaphragm and $\theta_{m}$ the lower cut-off angle determined by some
mask. If no mask is used, $\theta_{m}=0 \rightarrow q_{m}=0$.

Using the properties of Bessel functions Eq.(\ref{ib1}) can be developed as

\begin{eqnarray}
I(b) & = &4 \alpha \vert \int \int \frac{\mathrm{d}^2\mathbf{\vec{q}}}
{(2 \pi)^2}
\frac{\mathbf{\vec{q}}}{\mathbf{\vec{q}}^2+q_{0}^2} f(q)
e^{i \mathbf{\vec{q}} \cdot \mathbf{\vec{b}}} \vert^2 \nonumber \\
 & = & 4 \alpha \vert \mathbf{\vec{\nabla}_{\vec{b}}} \int \int 
\frac{\mathrm{d}^2\mathbf{\vec{q}}}{(2 \pi)^2}
\frac{f(q)}{\mathbf{\vec{q}}^2+q_{0}^2}
e^{i \mathbf{\vec{q}} \cdot \mathbf{\vec{b}}} \vert^2 \nonumber \\
 & = & \frac{\alpha}{\pi^2} \vert \mathbf{\vec{\nabla}_{\vec{b}}} 
\int q \mathrm{d}q \frac{f(q)}{q^2+q_{0}^{2}} J_{0}(q b) \vert^2
\nonumber \\
 & = & \frac{\alpha}{\pi^2} \vert \int q^2 \mathrm{d}q \frac{f(q)}
{q^2+q_{0}^{2}} J_{1}(q b)\vert^2 \label{ib2}
\end{eqnarray}

If we substitute for $f(q)$ the sharp cut-off function Eq.(\ref{fq}),
we obtain

\begin{equation}\label{ib3}
I(b) =  \frac{\alpha}{\pi^2} \vert \int_{q_{m}}^{q_{1}} \frac{q^2}
{q^2+q_{0}^{2}} J_{1}(q b) \mathrm{d}q \vert^2 
\end{equation}

This can be written using the angle $\theta =q\lambdabar$ 
and $q_{0}=\gamma^{-1}\lambdabar^{-1}$ (in natural units $\omega = 
\lambdabar^{-1}$) as

\begin{equation}\label{ib4}
I(b) =  C_{1}\vert \int_{\theta_{m}}^{\theta_{1}} 
\frac{\theta^2}{\theta^2+\gamma^{-2}} J_{1}\left(\frac{\theta}{\lambdabar}
b \right) \mathrm{d}\theta \vert^2
\end{equation}

If we rewrite the diffraction pattern given by Eq.(\ref{dpni}) using 
$a=f_{1}\theta$, $M$ = magnification = $f_{2}/f_{1}$ and the integration 
limits $\theta_{1}=a_{0}/f_{1}$ and $\theta_{m} = a_{m}/f_{1}$ (i.e.\ we 
have a mask), we obtain

\begin{equation}
I(\rho) = C_{2}\vert \int_{\theta_{m}}^{\theta_{1}} \frac{\theta^2}
{\gamma^{-2}+\theta^2} J_{1}\left(\frac{2 \pi}{\lambda} \frac{\theta}
{M}\rho \right) \mathrm{d} \theta \vert^2
\end{equation}

We can see that this is identical (excluding the constant factor) to 
Eq.(\ref{ib4}) taking into account the image magnification $\rho = Mb$.

\section{Summary and conclusions}

In this paper, we have considered the limitations brought by the diffraction
to the resolution of OTR images of high energy charged particles.
Starting from the scalar wave  theory, some basic formulas concerning 
the wave propagation in an optical system were recalled. Choosing, for 
the optical system, a telescope which exhibits very simple and interesting
properties, we have calculated the diffraction pattern of the OTR wave emitted 
by one electron. A virtual diaphragm, located in the common focal plane 
between the lenses of the telescope, allowed us to express the diffraction
in a rather simple way. The radial polarization of OTR was taken into account 
by considering the horizontal and vertical field components separately.

Our result coincides with that of V.A.Lebedev \cite{leb} and it is in 
qualitative agreement with that of D.W.Rule and R.B.Fiorito obtained in the 
scalar wave approximation \cite{rf}-\cite{rf2}. The obtained diffraction 
pattern was also compared to the well known standard diffraction pattern. 
The FWHM size of the  OTR diffraction pattern is by factor of $\sim$ 2.2 
wider than the "scalar OTR" pattern and by factor of $\sim$ 2.7 than the 
standard diffraction pattern.

Consideration of the general shape of the OTR diffraction pattern shows
that the FWHM width is insensitive for the particle energy,
whereas the tails increase with the energy. These tails may be seen
by very sensitive detectors: in that case, a central optical mask 
constitutes an effective cure.

In conclusion, up to energies considered ($\gamma \sim 5*10^4$),
the effects of the diffraction, evaluated by the FWHM of the OTR diffraction
pattern, are not limiting the resolution. The resolution depends more
likely on the properties of the experimental set-up, the contrast
sensitivity of the detector and the data treatment procedure.

\end{document}